# Spin Hall photoconductance in a 3D topological insulator at room temperature

**Paul Seifert[1], Kristina Vaklinova[2], Sergey Ganichev[3], Klaus Kern[2,4], Marko Burghard[2] and Alexander W. Holleitner[1,*]**

[1]*Walter Schottky Institut and Physik-Department, Technische Universität München, Am Coulombwall 4a, D-85748 Garching, Germany.*

[2]*Max-Planck-Institut für Festkörperforschung, Heisenbergstraße 1, D-70569 Stuttgart, Germany.*

[3]*Terahertz Center, University of Regensburg, D-93040 Regensburg, Germany.*

[4]*Institut de Physique, Ecole Polytechnique Fédérale de Lausanne, CH-1015 Lausanne, Switzerland.*

**Three-dimensional topological insulators are a class of Dirac materials, wherein strong spin-orbit coupling leads to two-dimensional surface states. The latter feature spin-momentum locking, i.e., each momentum vector is associated with a spin locked perpendicularly to it in the surface plane. While the principal spin generation capability of topological insulators is well established, comparatively little is known about the interaction of the spins with external stimuli like polarized light. We observe a helical, bias-dependent photoconductance at the lateral edges of topological $Bi_2Te_2Se$ platelets for perpendicular incidence of light. The same edges exhibit also a finite bias-dependent Kerr angle, indicative of spin accumulation induced by a transversal spin Hall effect in the bulk states of the $Bi_2Te_2Se$ platelets. A symmetry analysis shows that the helical photoconductance is distinct to common longitudinal photoconductance and photocurrent phenomena, but consistent with the accumulated spins being transported in the side facets of the platelets. Our findings demonstrate that spin effects in the facets of 3D topological insulators can be addressed and read-out in optoelectronic devices even at room temperatures.**

Selectively addressing the 2D surface states (SSs) of 3D topological insulators (TIs)[1-4] in electrical transport experiments is challenging due to the intrinsic doping of TIs, resulting in sizable contribution of the bulk states (BSs) to the conductivity. Nevertheless, electrically induced spin-polarized currents due to spin-momentum locking[5] have been demonstrated in lateral spin valve devices with a ferromagnetic spin detector based on a number of bismuth and antimony chalcogenides[6–9]. Alternatively, the SSs and their dynamics can be probed via optical excitation of photocurrents, whose decay enables access to the spin-relaxation time of both SSs and BSs in 3D TIs[10-15]. Time-resolved ARPES experiments have demonstrated dynamics control of spin-polarized currents in $Sb_2Te_3$[16,17] and $Bi_2Se_3$[11,18], as well as emergent Floquet-Bloch states due to hybridization of the SSs and pulsed circularly polarized light (CPL)[19]. According to theory[20,21], CPL can couple to the electron spin in a material and the helicity of the incident photons determines the spin-polarization of the resultant in-plane photocurrents due to asymmetric excitation of spin-up and spin-down carriers. This circular photogalvanic effect has been experimentally demonstrated in $Bi_2Se_3$[22–24], $Sb_2Te_3$[25], and $(Bi_{1-x}Sb_x)_2Te_3$[26] and can be tuned via electrostatic gating[23], Fermi energy tuning[26], and proximity interactions[27]. In addition, a helicity-dependent photovoltaic effect has been reported for normally incident light in vicinity of the metal contacts on a $Bi_2Se_3$ nanosheet[28].

Importantly, along with the SS-generated in-plane spin polarization, the strong spin-orbit coupling (SOC) of the bulk bands can give rise to out-of-plane spin polarization diffusing perpendicularly to the applied current direction via the intrinsic spin Hall effect[29-33]. Theory further predicts that a bulk-mediated diffusion of spin density can take place between the top and bottom TI surfaces[29]. So far, only little experimental data are available on bulk spin currents in 3D TIs. Spin-charge conversion via the inverse spin Hall effect has been demonstrated for a TI/metallic ferromagnet heterostructure[34]. In general, however, SO-enhanced scattering suggests strongly reduced spin diffusion lengths in 3D TIs, which makes spin detection and manipulation challenging. In the present work, we explore the spin texture of $Bi_2Te_2Se$ (BTS) platelets under electrical current flow using two complementary experimental methods, namely helical photoconductance measurements and magneto-optical Kerr microscopy. Our data is fully consistent with a bulk-spin Hall effect generating a spin accumulation in the side-facets of the BTS platelets, which can be read-out as a helical photoconductance with a lifetime characteristic of the SSs in the platelets' side-facets.

We contact individual BTS platelets on a transparent sapphire substrate by two Ti/Au contacts, and focus a circularly polarized laser with a photon energy of 1.54 eV normally incident onto the platelets (Fig. 1a and Methods). We apply a bias voltage $V_{sd}$ between the contacts, such that a local electron current density $j$ flows in the sample (black arrow in Fig. 1a). Figure 1b presents an optical microscope image of an exemplary device, comprising a BTS platelet with a width of 8.9 μm and a height of 90 nm, while the Au-contacts have a distance of 10 μm. We record the optical transmission signal of the platelet upon scanning the laser with a spot size of ~1.5 μm across (Fig. 1c ) and define three positions, specifically the left edge (blue circle), the center (gray circle), and the right edge (red circle). The corresponding photoconductance map acquired under an applied bias voltage $V_{sd}$ of 1.2 V (Fig. 1d) displays a helicity dependent photoconductance with opposite sign at both edges. In particular, these signals reflect the difference $\Delta G_{helical} = G_{helical}(\sigma^+) - G_{helical}(\sigma^-)$ of the photoconductance $G_{helical}$, which depends on the circularly right ($\sigma^+$) and right ($\sigma^-$) polarized light. The absence of a photoconductance signal away from the edges is consistent with the in-plane spin polarization associated to the SSs at the top/bottom surface of the platelet and the normal incidence of the circularly polarized light[17]. In order to gain further insight into the origin of the edge signals, we use a quarter wave plate to resolve the sign change of $G_{helical}$ as a function of helicity (Figs. 1e to g, and Supplementary Fig. 1). In the following, we interpret $G_{helical}$ to occur in the topological side surfaces of the BTS platelets as a consequence of spin-accumulation caused by the bulk spin-Hall effect[29-33]. In this scenario, the spin accumulation induced by the bias-driven current density $\boldsymbol{j}$ (black arrow in Fig. 1a) is characterized by an out-of-plane spin polarization pointing in opposite directions at the left and right edge of the BTS platelet (blue and red arrows).

Fig. 2a reveals that $\Delta G_{helical}$ depends linearly on the laser power $P_{laser}$ (up to 12 mW), without reaching saturation. By contrast, the polarization-independent photoconductance $G_{photo}$ displays a polarity change vs. laser power, as discussed in detail in [12]. Correspondingly, we attribute the polarity change to a saturation of the lifetime-limited photoconductance in the bulk states and the one in the SSs at the top and bottom surface for $P_{laser} \geq 5$ mW, while a bolometric photoconductance with a negative sign dominates the high power regime. We note that a bolometric photoconductance is driven by a heated phonon bath of the BTS platelets[12]. Fig. 2a highlights that $\Delta G_{helical}$ captures an effect that is clearly distinguished from the conventional longitudinal photoconductance effects in the bulk and surface states of BTS. This difference becomes particularly evident at a laser power of ~6 mW, where the polarization-independent $G_{photo}$ is zero, while $\Delta G_{helical}$ can be fully controlled with the laser polarization. The latter observation indicates that $\Delta G_{helical}$ is not a modulation of $G_{photo}$. To a first approximation, $\Delta G_{helical}$ is decoupled from the helicity independent longitudinal photoconductance, which justifies our approach of subtracting the helicity independent conductance contributions as an offset and only evaluating the helicity dependent photoconductance (cf. Supplementary Fig. 1).

The schematic sketches of the excitation configuration of our photoconductance experiment in Figs. 3a and b define the electron and photon wave vectors $k_{electron}$ and $k_{photon}$, as well as the spin polarization $\boldsymbol{S}$ at the left and right edges of the BTS platelet. As apparent from Figs. 3c to h, a finite $G_{helical}$ is observed only at a finite bias voltage, thereby excluding $G_{helical}$ as a photocurrent effect which would have a finite magnitude at zero bias. We note that for $V_{sd} = 0$ (Figs. 3d and g), the originally detected signal $I_{helical} \sim 0$ nA is plotted instead of $G_{helical}$, as an experimental conductance is not defined at zero bias. As another relevant observation, $G_{helical}$ flips by $\pi$ when either the bias $V_{sd}$ is reversed (cf. Fig. 3c vs. e and Fig. 3f vs. h), or when the laser spot position is exchanged between the left and right platelet edge.

Table 1 summarizes the experimentally observed symmetry characteristics of $G_{helical}$ as concluded from Figs. 1-3, and contrast them with those of the polarization-independent $G_{photo}$ and of helical photocurrents in materials with a broken symmetry[35,36]. The latter type of current is denoted as '$I_{edge}$' in the table, in order to associate it with the circular photogalvanic effect which is well documented for topological insulators[20-24,36]. From the observed bias dependence, any photocurrent effect can be excluded as origin of $G_{helical}$. However, it should be emphasized that such currents indeed emerge at the side-facets of the BTS platelets, although their amplitude is very small (cf. Supplementary Fig. S2). The small amplitude is explainable by the fact that the optical cross section of the side facets is about 100 times smaller compared to that of the bulk volume in the BTS platelets. It is furthermore noteworthy that the helical photoconductance $G_{helical}$ does not depend on the direction of the photon wavevector $k_{photon}$ (cf. Fig. 3 vs. Supplementary

Figs. 3 and 4). Moreover, from the polarity change of $G_{helical}$ with $V_{sd}$, one can exclude a longitudinal photoconductance response based upon the following argument. In case of a longitudinal response, the sign of $G_{helical}$ would reflect the preferred current direction in $x$-direction upon laser illumination. From a symmetry perspective, a current reversal is equivalent to a simultaneous operation 'left edge' to 'right edge' and a rotation of the sample by a spatial angle of $\pi$. The latter combined operations have to be an even operation due to rotational symmetry, while a bias reversal of $G_{helical}$ is odd (cf. table 1), which contradicts the experimental findings. Nonetheless, there remains a process which is fully consistent with the observed symmetries, specifically a transverse photoconductance. In fact, the latter can effectively modulate the conductivity along the edges of the BTS platelets, and thus supports the assumption of a bulk spin current in 3D TIs, which has been analytically predicted based upon topological arguments. Recently, bulk spin currents have been indentified as a possible spin generation mechanism in 3D topological insulators via the spin Hall effect[29-33]. Indeed, this mechanism would yield an accumulated spin polarization parallel to the SS-based spin polarization in the side facets of BTS platelets (cf. red and blue arrows in Fig. 1a), as recently verified and discussed[29-33,37]. Furthermore, it should result in an opposite spin accumulation at the opposite edges of the BTS platelets, and also reverse sign upon reversing the longitudinal charge current direction.

In order to verify the presence of vertically aligned spins accumulated at the edges of the BTS platelets, we investigate the device by magneto-optical Kerr-microscopy. The detected Kerr angles $\theta_k$ (Figs. 4a - c) are bias-dependent and their sign is opposite at the opposite edges of the platelet. Likewise, the sign of $\theta_k$ at the edges switches with the sign of the applied bias (Figs. 4a and c), whereas $\theta_k$ vanishes at the center of the platelet (Fig. 4b). In close correspondence, also $G_{helical}$ follows such bias and position dependence (Figs. 4d to f).

The above outlined mechanism of helicity dependent photoconductance in BTS can be rationalized by a microscopic model as follows. The spin relaxation of photogenerated charge and spin carriers in 3D TIs, including BTS, occurs on ultrafast timescales on the order of 100 fs[11]. This fast relaxation suggests that the underlying spin dynamics of $G_{helical}$ is governed by spin diffusion rather than a spin precession process[30]. The spin diffusion length in $Bi_2Se_3$ was experimentally determined to be below 10 nm at room temperature[38]. This length is significantly smaller than both, the diffusion length of hot carriers that ranges between 200 and 300 nm in bismuth-chalcogenides[39], and the laser spot size of ~1.5 μm in our experiment. Moreover, we observe that $G_{helical}$ is located at the edges of the BTS platelets (Fig. 1d). Interestingly, we can observe the edge-localized $G_{helical}$ also in a BTS platelet with a spatial width as small as $d_{platelet}$ ~ 300 nm (cf. Supplementary Fig. 5). The latter observation is consistent with a dominating spin diffusion length smaller than ½ · $d_{platelet}$ ~ 150 nm at the edges. It has been reported that the inverse Spin Hall effect in the bulk states of $n$-doped $Bi_2Se_3$ is comparable to or even dominating over the surface Edelstein effects[37, 38]. Hence, a bulk spin Hall effect should be favorable over an out-of-plan spin component due to warping effects in the top and bottom surface of the BTS platelets[20]. Moreover, the side facets are expected to exhibit a spin-momentum locking[40] even in the presence of disorder. Under sufficiently high surface disorder, the spin relaxation time has been predicted to be increased similar as the D'yakonov–Perel mechanism in a motional narrowing regime[29]. Analogously, the spin generation efficiency and surface conductivity are suggested to be strongly enhanced at the disordered facets of the BTS platelets[29], as previously demonstrated for surfaces of $Bi_1Sb_1TeSe_2$[41]. Along these lines, we interpret $G_{helical}$ as a measure of optically excited spins which are driven by the spin Hall effect to (or away from) the edges (blue and red arrows in Fig. 1a), where they experience an increased (or decreased) conductivity. This scenario is able to account for all our experimental findings. In particular, it explains not only the polarization dependence in Figs. 1e to g, but also all other experimental observations summarized in table 1. This interpretation suggests that $G_{\text{helical}}$ is a read-out of a bulk spin accumulation by scattering into the surface states of the side-facets in the BTS platetes[29]. Importantly, we determine the decay time of $G_{\text{helical}}$ to be $\tau_{\text{decay}} = (463 \pm 13)$ ps (cf. Supplementary Fig. 6). This value, while exceeding the spin relaxation time by far, agrees very well with the surface lifetime of the BTS platelets[12,42]; thereby strongly supporting the above mechanism. In other words, after a pulsed laser excitation, $G_{\text{helical}}$ prevails as long as $\tau_{\text{decay}}$ within the surface states of the side facets.

Overall, the possibility to optically detect the spin Hall effect-driven spins at structurally disordered 3D TI surfaces relaxes the technical demands on device fabrications, thus underscoring the theoretical prediction that surface disorder in 3D TIs might even be beneficial for spintronic applications[29]. Furthermore, the demonstrated read-out principle might prove useful also for probing the spatial accumulation of spins in related materials such as Weyl semimetals.

## Methods

**$Bi_2Te_2Se$ platelet growth and charge carrier density.** BTS platelets are grown by a catalyst-free vapor-solid method on Si substrates covered by a 300 nm thick $SiO_2$ layer. $Bi_2Se_3$ and $Bi_2Te_3$ crystal sources (99.999% purity) are heated up to ~580°C in a tube furnace. Ultrapure argon gas transports the evaporated material to the growth substrates, which are heated to 450°-480°C during a growth time of 30 min. The pressure in the furnace is maintained at 75-85 mbar at Argon flow rate of 150 sccm. Thus obtained platelets have a length in the range of 5-60 µm, a lateral width of 0.05-1 µm, and a thickness of 15-200 nm. The platelets are mechanically transferred onto $Al_2O_3$ substrates and lithographically contacted by metal strips made of thermally evaporated Ti/Au (10 nm/250 nm). Four-terminal Hall-measurements on single BTS platelets with a width of a few hundreds of nm revealed n-doping character, with an electron concentration of $10^{19}$ $cm^{-3}$ at room temperature.

**Time-integrated photoconductance spectroscopy.** For the excitation of the BTS platelets, we use a Ti-sapphire laser that emits either light at laser pulses with 150 fs pulse duration and a repetition rate of 76 MHz. In all cases, the photon energy is set to be 1.54 eV. The position of the laser spot with size ~1.5 µm is set with a spatial resolution of ~100 nm. The helical photoconductance is measured with a standard lock-in measurement technique, where the laser is modulated by a chopper at a frequency of $f_{chop}$ = 2.7 kHz or a piezoelastic modulator (PEM) at a frequency of 50 kHz. The PEM is set to switch the polarization of the exciting laser between $\sigma^-$ and $\sigma^+$ polarized light. Alternatively, we utilize a quarter wave plate to control the circular polarization of the laser. All presented experiments are performed at room temperature and in a vacuum chamber at ~5x$10^{-6}$ mbar.

**Autocorrelation photoconductance spectroscopy**. For characterizing the decay dynamics of $G_{helical}$, we split the pulsed laser beam into a pump and a probe pulse with comparable intensity, and focus the pump- (probe-) from the front (back) onto the BTS platelets (cf. Supplementary Fig. S4). The probe-beam, which is chopped at a frequency of ~1930 Hz, hits the sample at a time delay $\Delta t$. The time-resolved autocorrelation signal is then detected with a lock-in amplifier on the sum-fequency of the pump- and the probe-pulse as a function of the time delay $\Delta t$. The probe-pulse is delayed via a mechanical delay stage.

**Magneto-optical Kerr microscopy.** For the Kerr-effect spectroscopy we utilize the same measurement setup as for the photoconductance spectroscopy. The laser is polarized purely linearly. The polarization of the reflected light is rotated by 45° via a half-waveplate. Afterwards, a Wollaston prism splits the reflected light into two cross-polarized beams with an angle of 20° which are detected by two balanced photodetectors. The difference- and sum-signal of the detectors are detected by two lock-in amplifiers at a frequency of 2.7 kHz. To calibrate the reflected signal, the laser-spot is positioned onto a normally reflecting surface (e.g. gold) and the half-waveplate is adjusted, such that the difference signal of the detectors is perfectly zero. Any rotation of the reflected light's polarization shifts the balance between the detectors, which is then detected in the difference- but not in the sum-signal. Hence the sum-signal is used to calibrate the difference-signal in order to get rid of laser-intensity based noise and fluctuations.

## Acknowledgements

We acknowledge S. Tarasenko, H. Hübl, and M. Münzenberg for very helpful discussions, and the DFG (Projects HO 3324/8-2 and SFB 1277-A04) as well as the ERC-grant NanoREAL for financial support.

## Author contributions

P.S. and A.W.H. designed the experiments, K.V., M.B., K.K. provided the materials, P.S. performed all experiments, and P.S., A.W.H., S.G., K.V., K.K., M.B. analyzed the data and wrote the manuscript.

## Additional information

Supporting information is available online.

## Competing financial interests

The authors declare no competing interests.

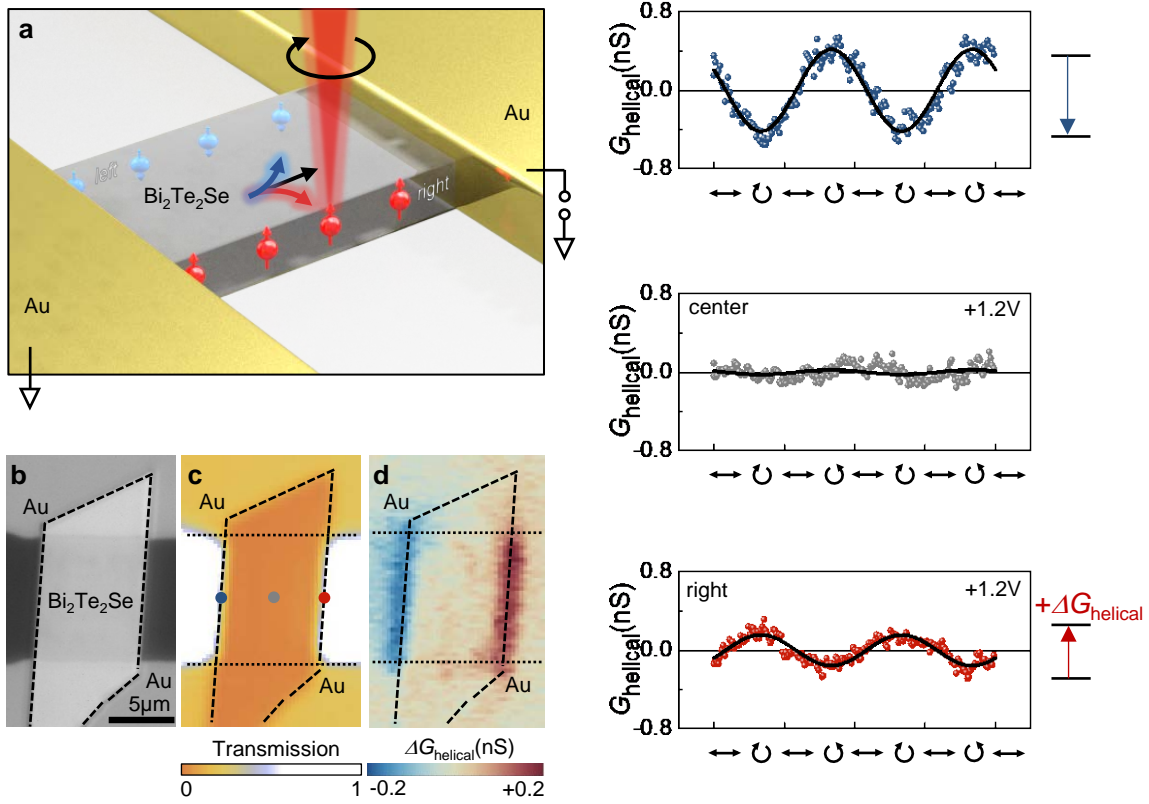

**Figure 1 | Helicity-dependent edge conductance at room temperature**. **a,** Sketch of a Bi₂Te₂Se (BTS) platelet embedded in a metal/topological-insulator/metal geometry excited by circularly polarized light at perpendicular incidence. Red (blue) spheres represent the electron spin polarization on the right (left) facet of the platelet, when a bias voltage $V_{sd}$ is applied between source to drain. Correspondingly, an electron current density $j$ flows. **b,** Optical microscope image of a BTS platelet (highlighted by dashed lines), contacted by two Ti/Au contacts. **c,** Optical transmission image of the platelet in (b), recorded with laser scanning microscopy. **d,** Simultaneously measured spatial map of the photoconductance generated by the circularly polarized light at a bias voltage $V_{sd} = 1.2$ V. We use a photo-elastic modulator detecting the difference $\Delta G_{helical} = G_{helical}(\sigma^-) - G_{helical}(\sigma^+)$ of photoconductance which depends on the circularly left ($\sigma^-$) and right ($\sigma^+$) polarized light. **e-g,** $G_{helical}$ as a function of the laser polarization at the left and right edge, as well as the center [cf. blue, gray, and red dots in (c)]. The polarization is controlled by a quarter waveplate. The symbol ↔ denotes linearly polarized, ↻ circularly right polarized light , ↺ circularly left polarized light. All measurements are performed at $V_{sd} = +1.2$ V, $P_{laser} = 230$ μW, and room temperature.

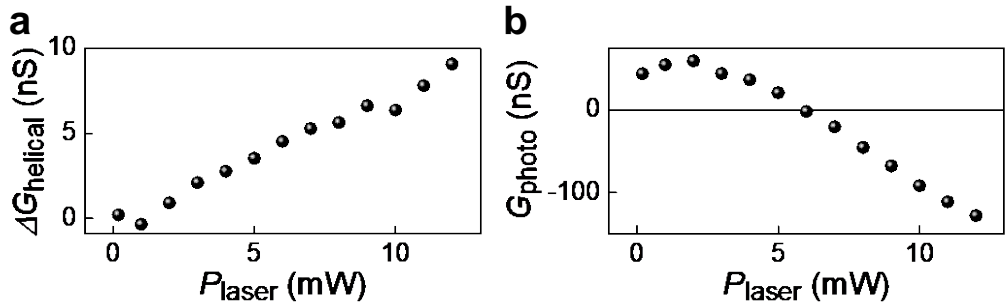

**Figure 2 | Irradiance dependence of helical and polarization-independent photoconductance. a,** The helical $\Delta G_{helical}$ is linear up to the largest laser powers investigated, and can be tuned to be positive or negative. The polarity depends on the helicity of the exciting photons and the direction of the dark current in the platelets. We show $\Delta G_{helical}$ for a right edge of a BTS platelet at a positive bias $V_{sd}$ = +0.55V (cf. Supplemental Figure S5). **b,** The polarization-independent photoconductance $G_{photo}$ of the BTS platelets shows a non-linear power dependence. The non-linearity is an interplay of saturating photoconductance contributions of the surface and bulk states of the BTS platelet (saturating for $P_{laser} \geq 5$ mW) and a negative bolometric photoconductance dominating at large laser powers [cf. ref.12]. All measurements are performed at room temperature.

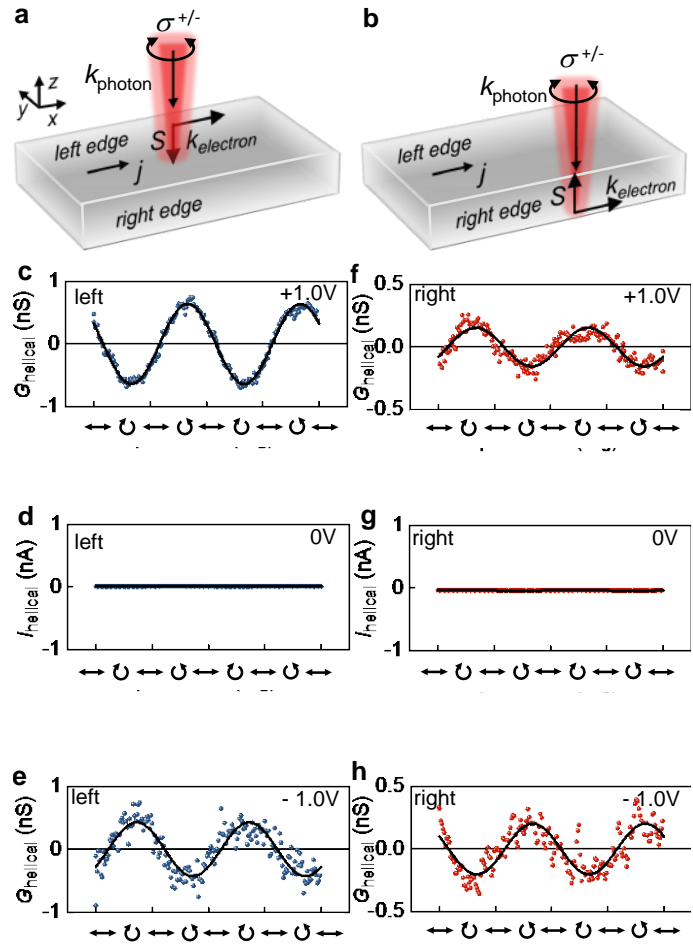

**Figure 3 | Helicity, bias, and facet dependence of helical photoconductance. a, b** Sketch of a BTS platelet with dark current *j* along the *x*-direction, when a bias $V_{sd}$ is applied. The current defines the direction of the electron wave vector $k_{electron}$ at the left and right edge. The electron spin-polarization **S** is perpendicular to $k_{electron}$ and points in opposite directions at the two edges. In (a) [(b)], the left [right] edge is excited with a circularly polarized light with the photon wave vector $k_{photon}$ pointing downwards. **c - e,** $\Delta G_{helical}$ vs the laser helicity at the left edge for $V_{sd} = +1$ V (c), 0 V (d), and -1 V (e). The polarization is tuned with a quarter waveplate for an excitation scheme as sketched in (a). **f - h,** Similar data for the right edge for an excitation scheme as sketched in (b). All measurements are performed at $P_{laser} = 230$ µW and room temperature.

| | $G_{\text{helical}}$ | $G_{\text{photo}}$ | $I_{\text{edge}}$ |
|---|---|---|---|
| $V_{\text{sd}}$ → $-V_{\text{sd}}$ | −1 | −1 | 0 |
| $\sigma^+$ → $\sigma^-$ | −1 | +1 | −1 |
| $k_{photon}$ → $-k_{\text{photon}}$ | +1 | +1 | +1 |
| *left edge* → *right edge* | −1 | +1 | −1 |

**Table 1 | Symmetry chart of $G_{\text{helical}}$, $G_{\text{photo}}$, and edge photocurrent phenomena.** Symmetry characteristics of $\Delta G_{\text{helical}}$, $G_{\text{photo}}$, and edge photocurrent effects $I_{\text{edge}}$ for changing $V_{\text{sd}}$ to $-V_{\text{sd}}$, $\sigma^+$ to $\sigma^-$, $k_{\text{photon}}$ to $-k_{\text{photon}}$, and the excitation position from the left to the right edge of the BTS platelets. A value '-1' states a sign change under the symmetry operation, while '+1' states a sign conservation. The value '0' denotes that $I_{\text{edge}}$ is independent from $j$ and the bias.

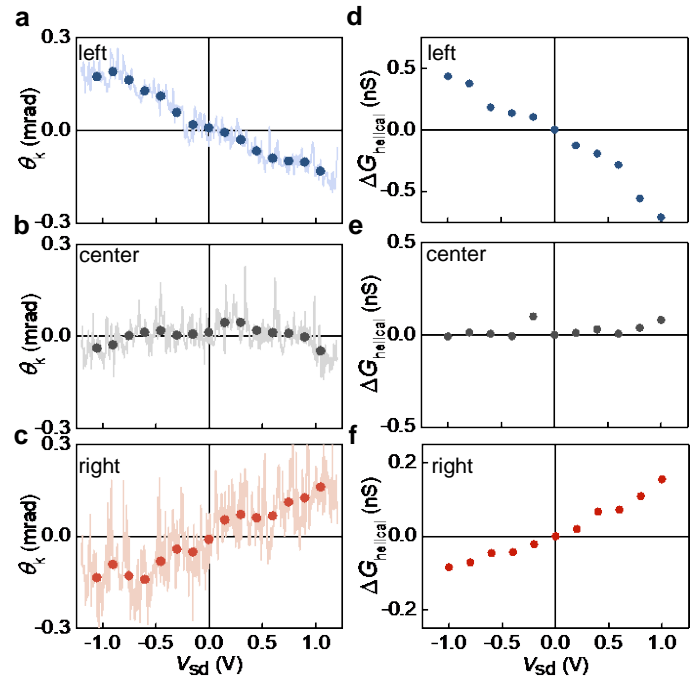

**Figure 4 | Kerr angle and helical photoconductance vs bias voltage. a - c,** Kerr angle $\theta_K$ vs the applied bias $V_{sd}$ at the left edge (a), center (b), and right edge (c) of a BTS platelet ($P_{laser} = 400\ \mu W$). **d - f,** Corresponding $\Delta G_{helical}$ vs the applied bias $V_{sd}$ ($P_{laser} = 230\ \mu W$). All measurements are performed at room temperature.

# Spin Hall photoconductance in a 3D topological insulator

## at room temperature

**Paul Seifert[1], Kristina Vaklinova[2], Sergey Ganichev[3], Klaus Kern[2,4], Marko Burghard[2] and Alexander W. Holleitner[1,*]**

[1]*Walter Schottky Institut and Physik-Department, Technische Universität München, Am Coulombwall 4a, D-85748 Garching, Germany.*

[2]*Max-Planck-Institut für Festkörperforschung, Heisenbergstraße 1, D-70569 Stuttgart, Germany.*

[3]*Terahertz Center, University of Regensburg, D-93040 Regensburg, Germany.*

[4]*Institut de Physique, Ecole Polytechnique Fédérale de Lausanne, CH-1015 Lausanne, Switzerland.*

## - Supplementary information -

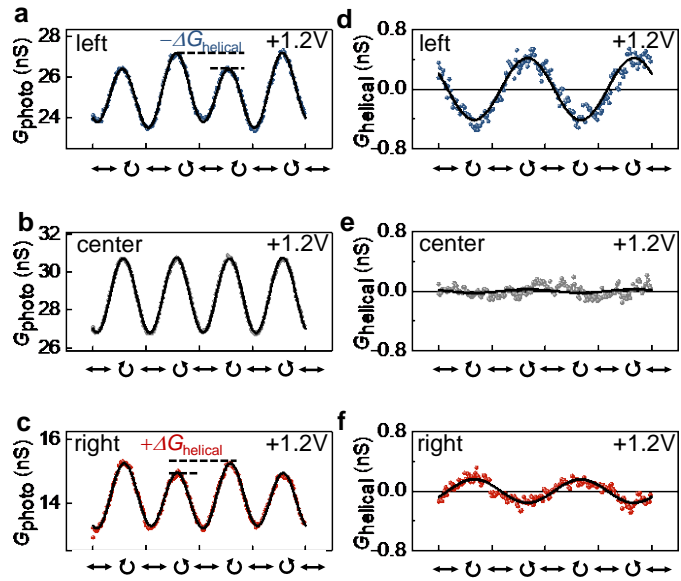

**Supplementary Figure S1 | Fitting procedure of the helical photoconductance. a, b, c,** Original data of the photoconductance vs the polarization of the exciting photons as depicted in Figs. 1e, 1f, and 1g of the main manuscript before substracting a background which depends on the linear polarization of the photon and a background which is independent of the polarization. The helicity-dependent contribution is indicated by $-\Delta G_{\text{helical}}$ and $+\Delta G_{\text{helical}}$. **d, e, f,** Helical photoconductance $G_{\text{helical}}$ which depends on the circularly right ($\sigma^+$) and left ($\sigma^-$) polarized light of (a)-(c) after substracting the background.

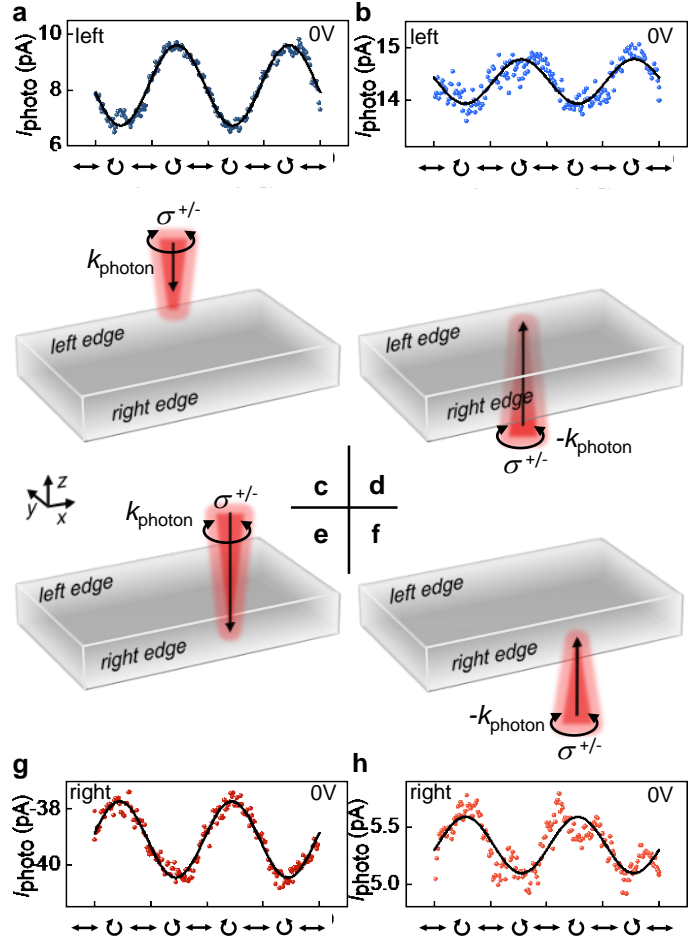

**Figure S2 | Helicity dependent edge *photocurrents* at zero bias. a,** and **b,** Photocurrent $I_{photo}$ at zero bias ($V_{sd} = 0$ V) for exciting the left side facet of the platelet as discussed in Figs. 1, 3, and 4 of the main manuscript, for a photoexcitation incident from the top (a) and backside (b) of the platelet. The polarization is controlled with a quarter waveplate. **c,** and **d,** Schematic illustrations for an excitation at the left side facet with a $\boldsymbol{k_{photon}}$ pointing downwards (c) and upwards (d). **e,** and **f,** Schematic illustrations for an excitation at the right side facet with a $\boldsymbol{k_{photon}}$ pointing downwards (e) and upwards (f). **g,** and **h,** $I_{photo}$ at zero bias ($V_{sd} = 0$ V) for exciting the right side facet of the platelet for a photoexcitation incident from the top (g) and from the backside (h). The polarization is controlled with a quarter waveplate.

Importantly, we note that for the side facets, the incidence occurs effectively at an oblique angle of ~90°. Hereby, photocurrents, such as the circular photogalvanic, are possible.[35] However, $I_{photo}$ exhibits a rather small amplitude of only pA, which is 1000x smaller than the $G_{helical}$ at $V_{sd} = 1$V (cf. Fig. 1). The small signal of $I_{photo}$ is consistent with the optical cross section of the side facets of only ~$10^{-2}$ of the overall laser spot of ~1.5 μm.

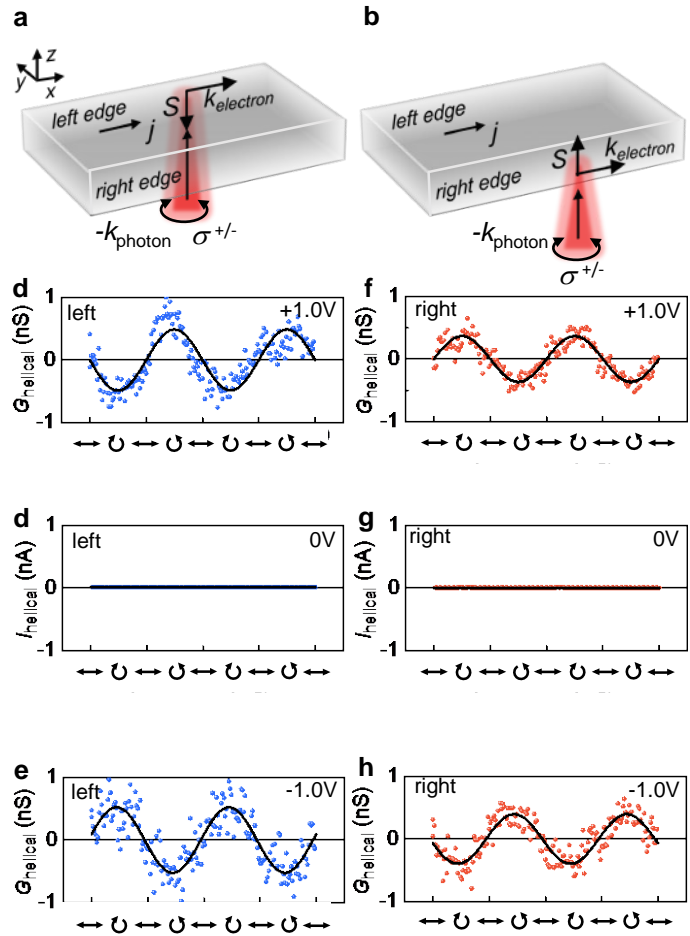

**Supplementary Figure S3 | Helical photoconductance at an opposite photon wave vector**. This graph shows supplementary data on the BTS platelet as discussed in Figs. 1, 3, and 4 of the main manuscript, but for a photo-excitation from the backside of the platelet. In other words, the photon wave vector is $-k_{photon}$ instead of $+k_{photon}$. **a,** and **b,** Schematic illustrations for the excitation configurations of the measurements in (c),(d),(e) and (f), (g), (h) respectively. **c, d, e,** Helical photoconductance $G_{helical}$ as a function of the laser polarization at the left edge (as depicted in (a)), for applied bias voltages of $V_{sd} = +1$ V, $V_{sd} = 0$ V and $V_{sd} = -1$ V. **f, g, h,** Helical photoconductance $G_{helical}$ as a function of the laser polarization at the right edge (as depicted in (b)), for applied bias voltages of $V_{sd} = +1$ V, $V_{sd} = 0$ V and $V_{sd} = -1$ V. We note that for $V_{sd} = 0$ (Supplementary Figs. S3d and S3g), we plot the originally detected signal $I_{helical} \sim 0$ nA instead of $G_{helical}$, because a photoconductance cannot be defined at zero bias. The polarization is controlled with a quarter waveplate.

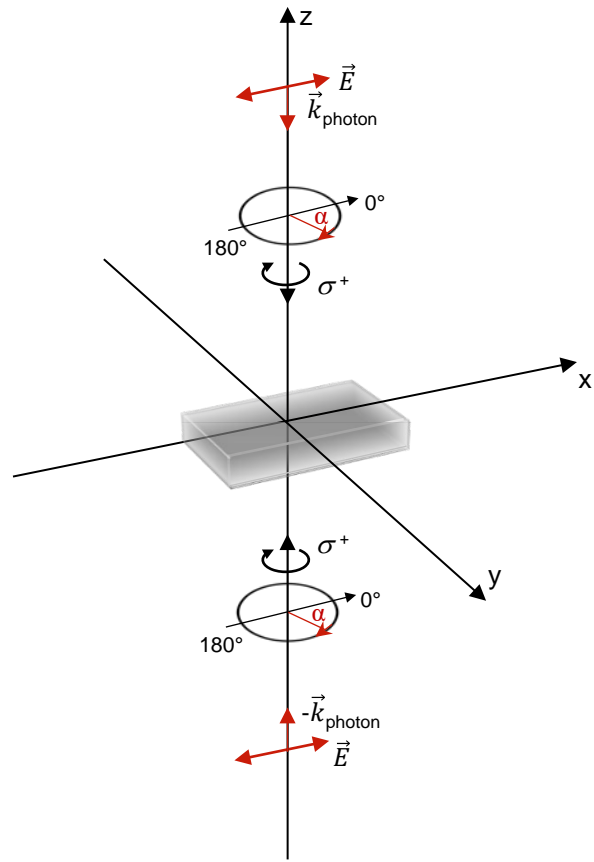

**Supplementary Figure S4 | Polarization and sample configuration.** Sketch of the measurement geometry and polarization optics circuitry. The quarter waveplates are arranged in a way, that the sense of rotation of the incoming laser light is defined globally and not with respect to the surface normal of the sample.

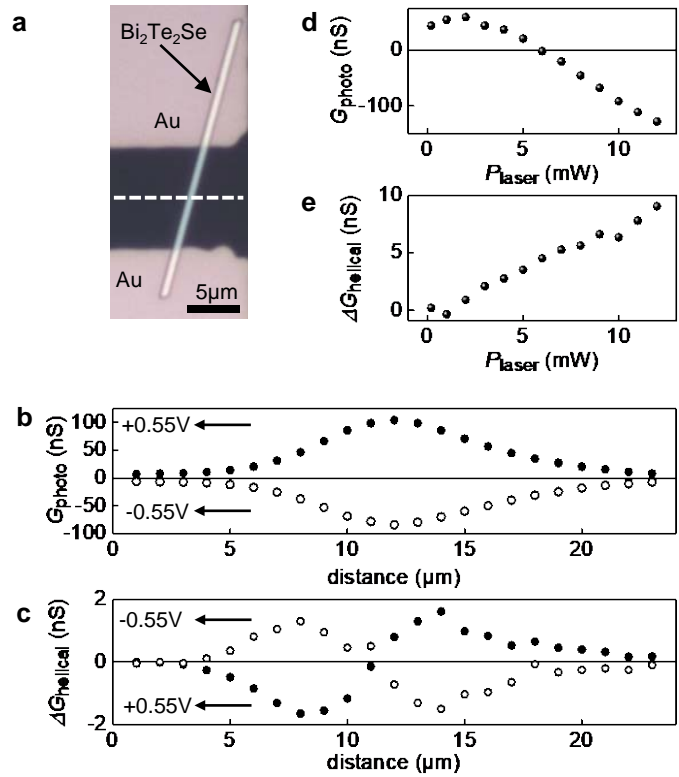

**Supplementary Figure S5 | Helical photoconductance in a narrow BTS platelet**. **a,** Optical microscope image of a narrow BTS nanoplatelet with a length of ~27 μm, a width of ~300 nm, and a height of 95 nm contacted by two Ti/Au contacts with a distance of 10 μm. This is sample of the data as discussed in Fig. 2 of the main manuscript. **b,** Total photoconductance $G_{photo}$ vs position along the white dashed line in (a) for $V_{sd}$ = +0.55 V (full circles) and -0.55 V (open circles). The measurements are performed at room temperature and $P_{laser}$ = 1.7 mW. **c,** Corresponding $\Delta G_{helical}$ along the white dashed line in (a). **d,** Photoconducance $G_{photo}$ as a function of incident laser power at $V_{sd}$ = +0.55 V. Same graph as Fig. 2b. **e,** Corresponding $\Delta G_{helical}$ as a function of incident laser power at $V_{sd}$ = +0.55 V. Same graph as Fig. 2a.

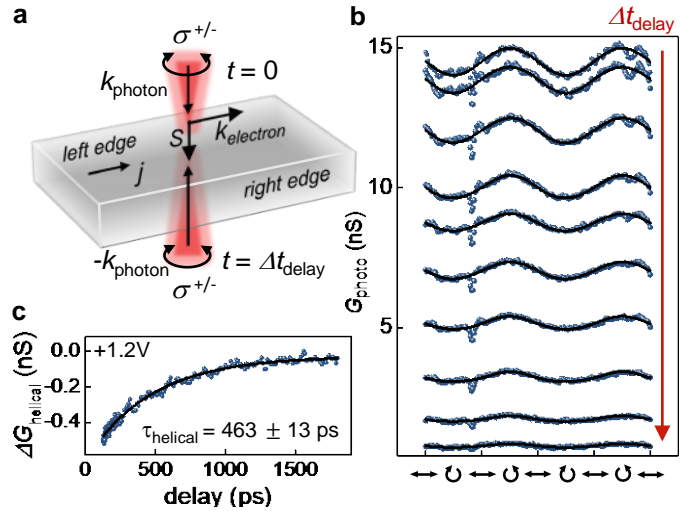

**Supplementary Figure S6 | Decay time of the helical photoconductance**. **a,** Schematic illustrations for measuring the auto-correlation of the helical photoconductance, when a finite bias $V_{sd}$ is applied, and correspondingly, an electron current density $j$ flows. **b,** Time-resolved pump/probe auto-correlation measurements at the left edge of a BTS platelet as a function of laser polarization. The top curve is recorded at a time-delay $\Delta t_{delay}$ = 0 ps between the pump and probe pulse. The bottom curve is recorded at $\Delta t_{delay}$ = 1800 ps. **c,** $\Delta G_{helical}$ as function of the time-delay $\Delta t_{delay}$ between pump- and probe-pulse. Measurement parameters are $P_{pump}$ = 1 mW, $P_{probe}$ = 2 mW, $V_{sd}$ = +1.2 V. The dimensions of the investigated BTS platelet are: length ~20 μm, width = 8.9 μm, height = 90 nm.